\documentclass[preprint]{elsarticle}

\usepackage{amsmath}    
\usepackage{amssymb}
\usepackage{graphicx}   
\usepackage{color}      
\usepackage{subfigure}  
\usepackage{hyperref}   

\newcommand{\bs}{\boldsymbol}

\begin{document}

\journal{Journal of Non-Newtonian Fluid Mechanics}

\title{Non-Newtonian effects on draw resonance\\in film casting}
\author{M. Bechert}
\ead{m.bechert@fz-juelich.de}
\address{Forschungszentrum J\"ulich GmbH, Helmholtz Institute Erlangen-N\"urnberg for Renewable Energy (IEK-11), F\"urther Stra\ss e 248, 90429 N\"urnberg, Germany}

\date{\today}

\begin{abstract}
In this paper, the influence of non-Newtonian material properties on the draw resonance instability in film casting is investigated. Viscoelastic models of infinite width film casting are derived systematically following an asymptotic expansion and using two well-known constitutive equations: the Giesekus model and the simplified Phan-Thien/Tanner~(PTT) model. Based on a steady state analysis, a numerical boundary condition for the inlet stresses is formulated, which suppresses the unknown deformation history of the die flow. The critical draw ratio in dependence of both the Deborah number and the nonlinear parameters is calculated by means of linear stability analysis. For both models, the most unstable instability mode may switch under variation of the control parameters, leading to a non-continuous change in the oscillation frequency at criticality. The effective elongational viscosity, which depends exclusively on the local Weissenberg number, is analyzed and identified as crucial quantity as long as the Deborah number is not too high. This is demonstrated by using a generalized Newtonian fluid model to approximate the PTT model. Based on such a generalized Newtonian fluid model, the effects of strain hardening and strain thinning are finally explored, revealing two opposing mechanisms underlying the non-Newtonian stability behavior.
\end{abstract}

\maketitle

\section{Introduction}
In film casting processes, molten material is extruded through a slit die, here referred to as the inlet, and taken-up by a distant, rotating chill roll, here referred to as the outlet (see Fig.~\ref{fig:sketch}(a)). Usually, the rotation speed at the outlet is desired to be high in order to produce films of preferably small thickness. The so-called draw resonance instability imposes an upper bound on the draw ratio, which is defined as the ratio of outlet to inlet velocity, and manifests itself in steady oscillations of flow velocity and both film thickness and width \citep{hatzikiriakos}. Many theoretical studies on this phenomenon were undertaken in the past decades, starting with a linear stability analysis of a purely viscous, one-dimensional model for infinite width film casting of a Newtonian fluid \citep{yeow74}. This model had been extended afterwards in several works to cover additional effects like inertia and gravity \citep{cao05,bechert15}, neck-in~\citep{silagy96,bechert16}, and cooling \citep{scheid09}.\\
In many cases, viscoelastic materials like polymer melts are processed and the non-Newtonian flow properties need to be taken into account in the modeling. First attempts were made employing a power-law equation in generalized Newtonian fluid (GNF) models \citep{pearson74, aird83,vanderhout00}. A power-law index larger than unity, i.e., pure strain hardening, was found to increase the critical draw ratio, while an index below unity, i.e., pure strain thinning, decreases the stability as compared to the well-known Newtonian value of $20.218$. \citet{silagy96} investigated the linear stability analysis of an upper-convected Maxwell~(UCM) fluid. A strongly stabilizing effect of increasing Deborah number is predicted, including a second critical draw ratio beyond which the process becomes stable again. Moreover, an unattainable region for high Deborah numbers and draw ratios is reported, where no steady state solutions exist.\\
Stability studies utilizing the Phan-Thien/Tanner~(PTT) \citep{lee07,shin07} and the eXtended Pom-Pom model \citep{vanderwalt12,gupta15} constitutive equations reveal in contrast to that a mostly destabilizing effect of increasing elastic properties of the material, which is in qualitative agreement with experimental findings \citep{zavinska08,burghelea12}. \citet{iyengar93,iyengar96} performed a steady state and a linear stability analysis of one-dimensional, infinite width film casting of a modified Giesekus fluid. The results are correlated with the dependence of the steady shear and uniaxial elongational viscosities on the deformation rates and shear thinning is found to lead to destabilization, while strain hardening can lead to a stabilization depending on the range of relaxation time and strain rates occurring within the process. While all of the above mentioned studies employ single mode models, there exist up to now also some works on draw resonance for a multi-relaxation-mode PTT model \citep{christodoulou00,dhadwal17,chougale18}. In two of these studies, the relaxation spectra are obtained from measurements of actual polymeric fluids \citep{christodoulou00,chougale18}.\\
Even though already many investigations on draw resonance can be found, including viscoelastic effects, some fundamental questions remain still unanswered. First of all, the influences of non-Newtonian, viscous effects and purely elastic effects on the critical draw ratio are not well separated yet. This includes the question, whether a GNF model is sufficient for essentially predicting the critical draw ratio or not. While \citet{iyengar96} attempted to investigate the stability mechanism of a deformation dependent viscosity, their analysis suffers from two major drawbacks: They used the uniaxial elongational viscosity, while the one-dimensional film casting process is governed exclusively by planar deformation, and they neglected the time dependence of the viscosity by accounting for the steady values, which are valid only for sufficiently long lasting deformations at constant deformation rate. For these reasons, it is still unclear if strain hardening (strain thinning) always leads to stabilization (destabilization) and why (not). Moreover, the qualitative difference between the stabilizing influence of elasticity as predicted by the UCM model on the one side, and the opposite predictions of other viscoelastic models on the other side has not been explained up to now, which is strongly connected to the minimum requirements for a constitutive equation to yield at least qualitative coincidence with experimental findings. The present study aims to address these questions, as well as to provide a comprehensive and consistent picture of the stability behavior of the commonly most used viscoelastic models.\\
The paper is organized as follows. We start with the derivation of the one-dimensional models following an asymptotic expansion of the flow equations in Sec.~\ref{sec:model}. We will use a uniform notation which enables us to use the same equations for both employed constitutive equations by simply tuning the corresponding parameters. Initially, an empirical stress boundary condition will be defined. Then, a steady state analysis is performed in Sec.~\ref{sec:steady}, which primarily serves the purpose of refining the stress boundary condition in a more consistent, physical way. Afterwards, the linear stability equations are derived and the stability results are presented in Sec.~\ref{sec:stability} for both constitutive equations. We then focus on non-Newtonian, viscous models in Sec.~\ref{sec:GNF} and analyze in particular the effective elongational viscosity. Based on this quantity, we test the possibility to approximate the previously studied viscoelastic models by GNF models, and discuss the limitations of this approach. Finally, the mechanisms underlying the stability behavior of strain hardening and strain thinning materials are revealed in Sec.~\ref{sec:mech}. The paper closes with brief conclusions and outlook.

\section{The model}
\label{sec:model}
\subsection{Governing equations}
\begin{figure}
\centering
\includegraphics[width=\textwidth]{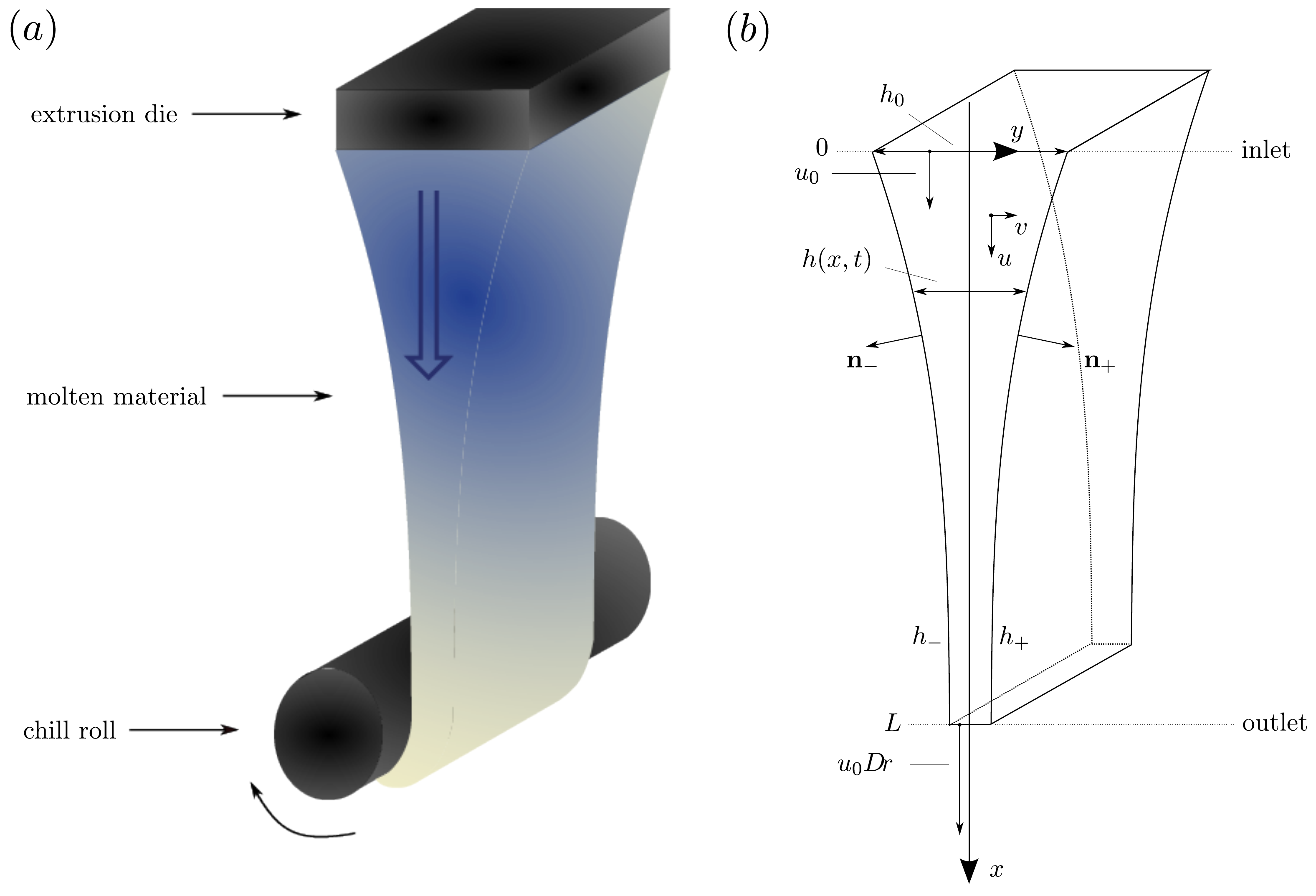}
\caption{(a) Sketch of the film casting process. The flow direction is indicated with a big arrow. (b) Sketch of the one-dimensional model for an infinitely wide film.}
\label{fig:sketch}
\end{figure}

The infinite width film casting model as presented by \citet{yeow74} is extended to viscoelastic constitutive equations. For this purpose, an asymptotic expansion in a small film parameter is employed to obtain the one-dimensional equations at leading order, following previous studies on Newtonian fluids \citep{bechert16,scheid09,eggers08}. Figure~\ref{fig:sketch}(b) shows a sketch of the film model between inlet and outlet. Assuming an infinite width of the film, it is sufficient to analyze the cross section of the film in the $xy$-plane, which is described by the film thickness $h(x,t)$ and the casting length $L$. The velocity field is denoted by $\bs v = (u,v)$, with axial velocity $u$ and transversal velocity $v$, and $t$ denotes the time. At the inlet, axial velocity and thickness are prescribed by the extrusion to $u_0$ and $h_0$, respectively. At the outlet, the axial velocity is fixed by the take-up with the chill roll to $u_L=D\!r\,u_0$, where the draw ratio $D\!r$ is introduced as primary control parameter.\\
The continuity and momentum conservation equations, given by, respectively,
\begin{subequations}
\begin{align}
\label{eq:cont_gen}
\bs\nabla\cdot\bs v &= 0,\\
\label{eq:stokes_gen}
\bs\nabla\cdot\bs\sigma &= 0,
\end{align}
\end{subequations}
with $\bs\nabla = (\partial_x,\partial_y)$ and stress tensor $\bs\sigma$, are coupled to the general constitutive equation
\begin{align}
\label{eq:const_eq}
\lambda\,\overset{\nabla}{\bs{\tau}}+\bs Y(\bs\tau) = \eta_0\,\left(\bs\nabla\bs v + (\bs\nabla\bs v)^T\right),
\end{align}
with upper-convected time derivative
\begin{align}
\overset{\nabla}{\bs{\tau}} = \partial_t\bs\tau + \bs v\cdot\bs\nabla\bs\tau - \bs\tau\cdot\bs\nabla\bs v - (\bs\nabla\bs v)^T\cdot\bs\tau
\end{align}
of the extra stress tensor $\bs\tau=p\bs 1 + \bs\sigma$, $p$ denoting the pressure. $\lambda$ is the relaxation time and $\eta_0$ the zero-shear viscosity, both assumed to be constant. The auxiliary tensor $\bs Y$ is defined as
\begin{align}
\bs Y(\bs\tau) &= \begin{cases}
\bs \tau + \alpha \frac{\lambda}{\eta_0}\bs\tau^2\ &\text{for the Giesekus model},\\[10pt]
\exp\left(\varepsilon\frac{\lambda}{\eta_0}tr(\bs\tau)\right)\bs\tau &\text{for the PTT model},
\end{cases}
\end{align}
$tr(\bs\tau)$ denoting the trace of $\bs\tau$.
The Giesekus and PTT models \citep{larson} introduce an additional, nonlinear parameter denoted by $\alpha$ or, respectively, $\varepsilon$. Note that only the so-called simplified PTT model with one nonlinear parameter is treated in this paper. In the limit cases of $\alpha\rightarrow0$ or $\varepsilon\rightarrow0$, the UCM model is recovered. In the following, we will shorten the notation by combining the two models with the note that one of the nonlinear parameters, $\alpha$ or $\varepsilon$, has to be set to zero depending on the chosen model, i.e., 
\begin{align}
\bs Y(\bs\tau) = \exp\left(\varepsilon\,\lambda/\eta_0\,tr(\bs\tau)\right)\bs\tau+\alpha\,\lambda/\eta_0\,\bs\tau^2.
\end{align}
At the free surfaces $h_{\pm}$ defined by $y=\pm h/2$, the kinematic and stress boundary conditions,
\begin{subequations}
\begin{align}
\label{eq:kin_gen}
\partial_t h_\pm + u|_{h_\pm}\partial_xh_\pm-v|_{h_\pm} &= 0,\\
\label{eq:str_gen}
\bs n_\pm \cdot \bs\sigma |_{h_\pm} &= 0,
\end{align}
with normal vector $\bs n_\pm = (-\partial_xh_\pm,1)/\sqrt{1+(\partial_xh)^2/4}$, are imposed.
\end{subequations}
\subsection{Scaling and asymptotic expansion}
The system is scaled according to the following transformations,
\begin{align}
\begin{split}
x \rightarrow L\,x ,\ \ y \rightarrow \beta\, L\, y ,\ \ u \rightarrow u_0\, u ,\ \ v \rightarrow \beta\, u_0\, v ,\\h \rightarrow \beta\, L\, h ,\ \ t \rightarrow \frac{L}{u_0}\, t ,\ \ \bs\tau\rightarrow \eta\,\frac{u_0}{L}\, \bs\tau,\ \ p \rightarrow \eta\,\frac{u_0}{L}\, p,\ \ \bs\sigma\rightarrow \eta\,\frac{u_0}{L}\, \bs\sigma,
\end{split}
\end{align}
with the film parameter $\beta=h_0/L$ assumed to be small. From now on, only scaled variables will be used unless explicitly stated differently.\\
The continuity equation~(\ref{eq:cont_gen}) can then be written,
\begin{align}
\label{eq:cont}
\partial_x u + \partial_y v &= 0,
\end{align}
and the two components of the momentum conservation~(\ref{eq:stokes_gen}) become
\begin{subequations}
\begin{align}
\label{eq:stokes_x}
\beta^2\,\partial_x\sigma_{xx}+\partial_y\sigma_{xy}^* &= 0,\\
\label{eq:stokes_y}
\partial_x\sigma_{xy}^*+\partial_y\sigma_{yy} &= 0,
\end{align}
\end{subequations}
with $\sigma_{xy}^*=\beta\,\sigma_{xy}$. The components of the stress boundary condition~(\ref{eq:str_gen}) read
\begin{subequations}
\begin{align}
\label{eq:str_x}
-\beta^2\,\partial_xh_\pm\,\sigma_{xx}|_{h_\pm} + \sigma_{xy}^*|_{h_\pm} &= 0,\\
\label{eq:str_y}
-\partial_xh_\pm\,\sigma_{xy}^*|_{h\pm} + \sigma_{yy}|_{h_\pm} &= 0,
\end{align}
\end{subequations}
and the constitutive equation~(\ref{eq:const_eq}) is given by
\begin{subequations}
\begin{align}
\label{eq:const_xx}
\begin{split}
D\!e\left(\beta^2\left(\partial_t\tau_{xx}+u\,\partial_x\tau_{xx}+v\,\partial_y\tau_{xx}\right)\right.\\\left.-2\left(\beta^2\,\tau_{xx}\,\partial_xu+\tau_{xy}^*\,\partial_yu\right)\right) + Y_{xx} &= 2\,\beta^2\,\partial_xu,
\end{split}
\\
\label{eq:const_yy}
\begin{split}
\beta^2\,D\!e\left(\partial_t\tau_{yy}+u\,\partial_x\tau_{yy}+v\,\partial_y\tau_{yy}-\right.\\\left.2\left(\tau_{yy}\,\partial_yv+\tau_{xy}^*\,\partial_xv\right)\right) + Y_{yy} &= 2\,\beta^2\,\partial_yv,
\end{split}
\\
\label{eq:const_xy}
\begin{split}
D\!e\left(\partial_t\tau_{xy}^*+u\,\partial_x\tau_{xy}^*+v\,\partial_y\tau_{xy}^*-\tau_{xy}^*\,\partial_yv\right.\\\left.-\beta^2\,\tau_{xx}\,\partial_xv-\tau_{xy}^*\,\partial_xu-\tau_{yy}\,\partial_yu\right) + Y_{xy} &= \beta^2\,\partial_xv+\partial_yu,
\end{split}
\end{align}
\end{subequations}
where $\tau_{xy}^* = \beta\,\tau_{xy}$, and with Deborah number $D\!e = \lambda\,u_0/L$ and
\begin{subequations}
\begin{align}
Y_{xx} &= \beta^2\,\tau_{xx}\exp\left[\varepsilon\,D\!e(\tau_{xx}+\tau_{yy})\right] + \alpha\,D\!e\left(\beta^2\,\tau_{xx}^2+\tau_{xy}^{*2}\right),\\
Y_{yy} &= \beta^2\,\tau_{yy}\exp\left[\varepsilon\, D\!e\, (\tau_{xx}+\tau_{yy})\right]+\alpha\,D\!e\left(\beta^2\,\tau_{xx}^2+\tau_{xy}^{*2}\right),\\
Y_{xy} &= \tau_{xy}^*\exp\left[\varepsilon\, D\!e\, (\tau_{xx}+\tau_{yy})\right]+ \alpha\,D\!e\left(\beta^2\,\tau_{yy}^2+\tau_{xy}^{*2}\right).
\end{align}	
\end{subequations}
The stresses, pressure, and velocity variables are now expanded in the square film parameter $\beta^2$, i.e., for any variable $\phi \in \{u,v,p,\bs\tau,\bs\sigma\}$,
\begin{align}
\phi = \phi^{(0)} + \beta^2\,\phi^{(1)} + \mathcal{O}(\beta^4).
\end{align}
Equations~(\ref{eq:stokes_x}) and (\ref{eq:str_x}) truncated at $\mathcal{O}(\beta^2)$ then lead to $\sigma_{xy}^{*(0)}=\tau_{xy}^{*(0)}=0$, and therefore Eq.~(\ref{eq:const_xy}) yields
\begin{align}
\label{eq:visc_cond}
\left(D\!e\,\tau_{yy}^{(0)}+1\right)\partial_yu^{(0)} = 0
\end{align}
at leading order. Assuming $\partial_y u^{(0)}$ being non-zero would directly lead to
\begin{align}
\tau_{yy}^{(0)} = -\frac{1}{D\!e},
\end{align}
which implies that the transversal stress component diverges in the case of the Newtonian limit $D\!e\rightarrow0$. As this is obviously not the case, the only consistent way to fulfill Eq.~(\ref{eq:visc_cond}) is $\partial_yu^{(0)}=0$, i.e., the axial velocity depends only on $x$ and $t$ at leading order.

\subsection{Averaged equations}
Averaging the continuity equation~(\ref{eq:cont}) at leading order over the thickness $h$, together with employing the kinematic boundary condition\footnote{The kinematic boundary condition remains unchanged by the scaling transformation.} (\ref{eq:kin_gen}), we obtain the one-dimensional form of the continuity equation,
\begin{align}
\label{eq:1D_cont}
\partial_t h + \partial_x\left(h\,u^{(0)}\right) &= 0.
\end{align}
%
%
Equations~(\ref{eq:stokes_y}) and (\ref{eq:str_y}) yield $\sigma_{yy}^{(0)}=0$, which makes $\sigma_{xx}^{(0)}$ equal to the normal stress difference $\nu^{(0)}=\tau_{xx}^{(0)}-\tau_{yy}^{(0)}$ at leading order. Averaging the $x$-component of the momentum equation~(\ref{eq:stokes_x}) over the film thickness by utilizing the first order solution at the boundaries $\sigma_{xy}^{*(1)}$ as obtained from Eq.~(\ref{eq:str_x}) finally leads therefore to
\begin{align}
\label{eq:1D_mom}
\partial_x\left(h\,\overline{\nu^{(0)}}\right) = 0,
\end{align}
with the depth-averaged normal stress difference at leading order given by
\begin{align}
\overline{\nu^{(0)}} = \frac{1}{h}\int_{-h/2}^{h/2}dy\,\nu^{(0)}.
\end{align}
Finally, the first non-vanishing order of Eqs.~(\ref{eq:const_xx}) and (\ref{eq:const_yy}) is averaged along the film thickness. For this we have to assume additionally
\begin{align}
\left|\tau_{ii}^{(0)} - \overline{\tau_{ii}^{(0)}}\right| = \mathcal{O}(\beta),
\end{align}
as this ensures that we can calculate the average of nonlinear stress terms at leading order by
\begin{align}
\label{eq:av_nonlinear}
\frac{1}{h}\int_{-h/2}^{h/2}dy\,\left(\tau_{ii}^{(0)}\right)^2 = \left(\overline{\tau_{ii}^{(0)}}\right)^2.
\end{align}
Employing the continuity equation (\ref{eq:cont}), one then obtains
\begin{subequations}
\label{eq:1D_const}
\begin{align}
\begin{split}
D\!e\left[\partial_t\overline{\nu^{(0)}}+u^{(0)}\partial_x\overline{\nu^{(0)}}+2\,\partial_xu^{(0)}\left(\overline{\nu^{(0)}}-2\,\overline{\tau_{xx}^{(0)}}\right)\right]\\ + \,Z_{\nu}\,\overline{\nu^{(0)}} &= 4\,\partial_xu^{(0)},
\end{split}\\
\begin{split}
D\!e\left(\partial_t\overline{\tau_{xx}^{(0)}}+u^{(0)}\partial_x\overline{\tau_{xx}^{(0)}}-2\,\partial_xu^{(0)}\overline{\tau_{xx}^{(0)}}\right)\\ + \,Z_{xx}\,\overline{\tau_{xx}^{(0)}} &= 2\,\partial_xu^{(0)},
\end{split}
\end{align}
\end{subequations}
with
\begin{subequations}
\begin{align}
Z_{\nu} &= \exp\left[\varepsilon\,D\!e\,\left(2\,\overline{\tau_{xx}^{(0)}}-\overline{\nu^{(0)}}\right)\right] + \alpha\,D\!e\,\left(2\,\overline{\tau_{xx}^{(0)}}-\overline{\nu^{(0)}}\right),\\
Z_{xx} &= \exp\left[\varepsilon\,D\!e\,\left(2\,\overline{\tau_{xx}^{(0)}}-\overline{\nu^{(0)}}\right)\right] +\alpha\,D\!e\,\overline{\tau_{xx}^{(0)}},
\end{align}
\end{subequations}
where a generalized version of Eq.~(\ref{eq:av_nonlinear}) was used to average the exponential term of the PTT model.\\
Equations~(\ref{eq:1D_cont}), (\ref{eq:1D_mom}) and (\ref{eq:1D_const}) determine the evolution of the four state variables $u^{(0)}$, $h^{(0)}$, $\overline{\nu^{(0)}}$ and $\overline{\tau_{xx}^{(0)}}$. We will omit the averaging bar and the superscript ``$(0)$'' from now on and simply use $u$, $h$, $\nu$ and $\tau_{xx}$ for the variables at leading order. Nevertheless, it is important to point out that, in contrast to the Newtonian model, the stress components are not necessarily identical to their averaged quantities. This is usually ignored in the present literature and the equations are mostly derived by assuming \textit{a-priori} transversal invariance of the variables as well as neglecting off-diagonal stress components. 

\subsection{Boundary conditions (empirical approach)}
\begin{figure}[t]
\centering
\includegraphics[width=0.7\textwidth]{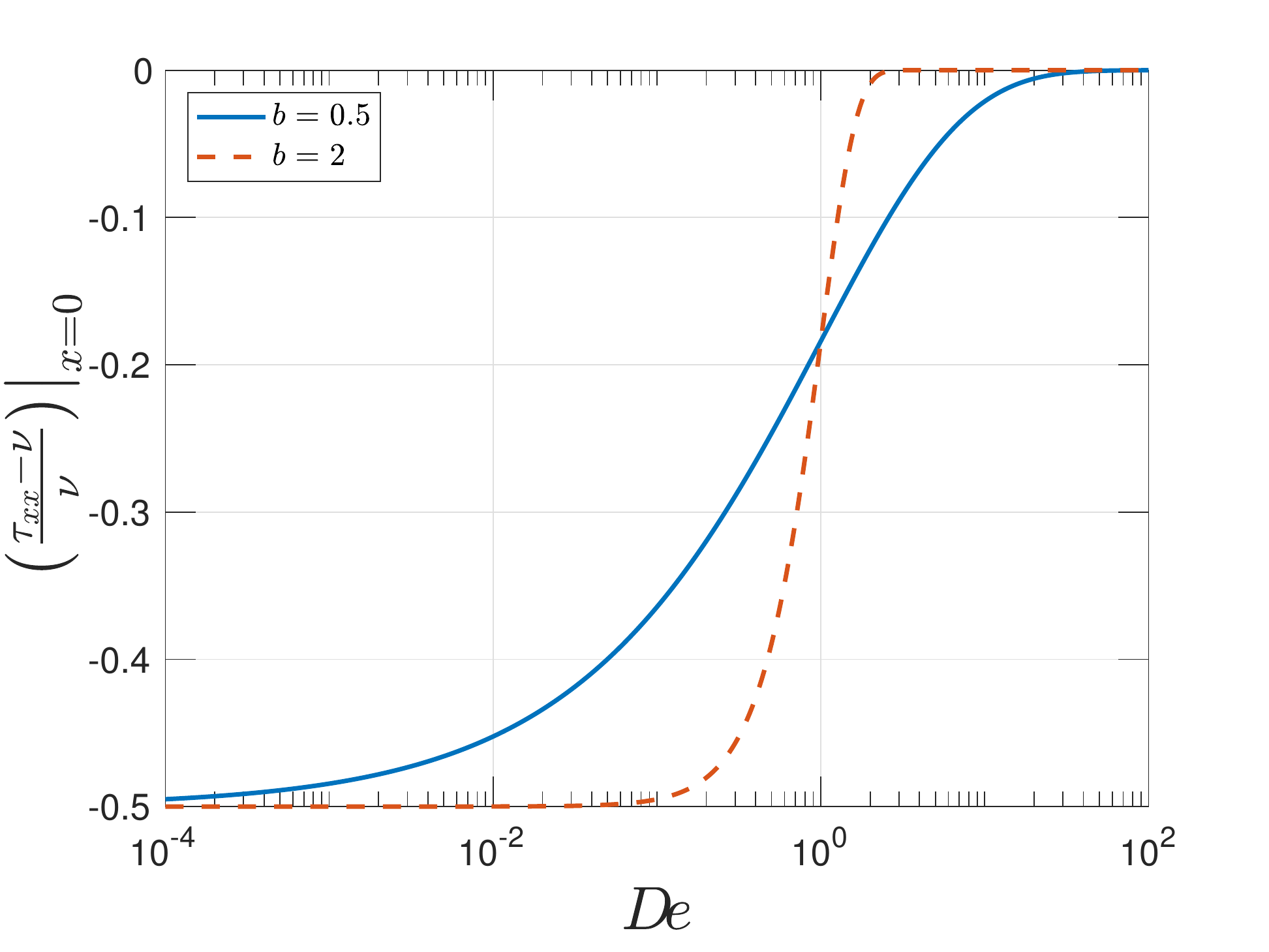}
\caption{Empirical stress boundary condition (\ref{eq:str_bc_naive}) for $b=0.5$ and $2$.}
\label{fig:BC_naive}
\end{figure}

Three boundary conditions are given directly by the process setup,
\begin{subequations}
\label{eq:bc}
\begin{align}
u(x=0,t) = h(x=0,t) &= 1,\\
u(x=1,t) &= D\!r.
\end{align}
\end{subequations}
With four first-order differential equations~(\ref{eq:1D_cont}), (\ref{eq:1D_mom}) and (\ref{eq:1D_const}), a fourth spatial boundary condition is needed. This one is physically determined by the pre-history of the flow in the die, which defines the stress components at the inlet. A comprehensive analysis including the die flow increases significantly the complexity of the problem, which is why we will follow an alternative approach here. As already pointed out by \citet{papanastasiou_96} for fiber spinning, the ratio of transversal stress to the normal stress difference decreases monotonically with increasing elastic effects. At the Newtonian limit, i.e., $D\!e\rightarrow0$, the value is fixed to $\tau_{yy}/\nu = -1/2$ and in the elastic limit, i.e., $D\!e\rightarrow\infty$, the transversal stress vanishes. \citet{papanastasiou_96} proposed a so-called free boundary condition method to define implicitly a stress inlet condition using finite element methods. This boundary condition is equivalent to assuming that the deformation in the die is exactly the same as outside the die, i.e., purely extensional, and leads therefore to smooth stress profiles without boundary layers at the inlet. Here, we will introduce a rather empirical, but more flexible approach, which will be elaborated completely in Sec.~\ref{sec:eff_visc}. For now, a qualitative expression based on the physical considerations above is used to obtain first results,
\begin{align}
\label{eq:str_bc_naive}
\left.\left(\frac{\tau_{xx}-\nu}{\nu}\right)\right |_{x=0} = -\frac{1}{2\,}\exp\left({-\,D\!e^b}\right),
\end{align}
which tunes the transversal stress $\tau_{yy}=\tau_{xx}-\nu$ at the inlet exponentially from the Newtonian limit to zero with increasing Deborah number. The transition can be adjusted by parameter~$b$, as it is depicted in Fig.~\ref{fig:BC_naive} for two values of $b$.
\section{Steady state analysis}
\label{sec:steady}
\subsection{Steady state equations}
The steady state equations, which determine the time-independent solutions, can be obtained by neglecting the time derivatives in Eqs.~(\ref{eq:1D_cont}), (\ref{eq:1D_mom}) and (\ref{eq:1D_const}). This leads to the following system,
\begin{subequations}
\label{eq:steady}
\begin{align}
u_s' &= \frac{\nu_s}{\tilde\mu_s},\\
h_s' &= -\frac{h_s\nu_s}{u_s\,\tilde\mu_s},\\
\nu_s' &= \nu_s\frac{u_s'}{u_s},\\
t_s' &= \frac{1}{D\!e\,u_s}\left[2\,u_s'(1+D\!e\,t_s)-t_s\,Z_{xx,s}\right],
\end{align}
\end{subequations}
where we introduce the \textit{effective elongational viscosity} at steady state,
\begin{align}
\label{eq:eff_visc}
\tilde\mu_s = \frac{\nu_s}{u_s'} = \frac{4}{Z_{\nu,s}}\,\left[1 + D\!e\left(t_s-\frac{3}{4}\,\nu_s\right)\right],
\end{align}
which will be investigated in more detail in Sec.~\ref{sec:GNF}. The steady state variables are indicated by a subscript `$s$' and $t_s=\tau_{xx,s}$ is used for the sake of convenience. Equations~(\ref{eq:steady}) are solved by numerical continuation employing the software \textsc{auto-07p}\citep{auto07p}, using the Newtonian limit as an initial solution.
\begin{figure}
\centering
\includegraphics[width=.8\textwidth]{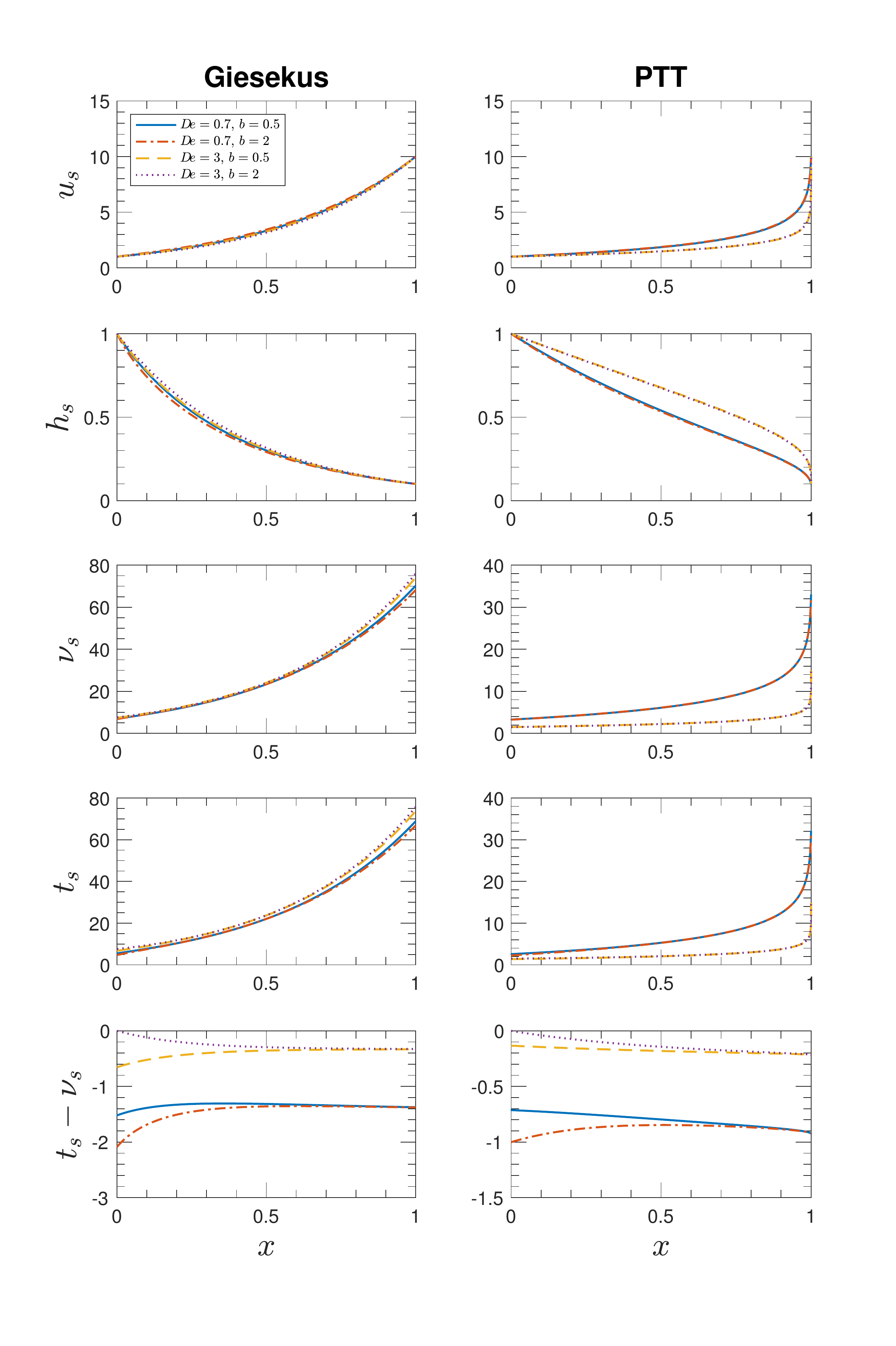}
\caption{Steady state solutions for the Giesekus and PTT models employing the empirical inlet stress boundary condition (\ref{eq:str_bc_naive}) for $b=0.5$ and $2$ at two Deborah numbers $D\!e=0.7$~and~$3$, with $D\!r=10$. For the Giesekus model $\alpha=0.3$ and for the PTT model $\varepsilon=0.3$.}
\label{fig:steady_states}
\end{figure}

\subsection{Influence of stress boundary condition}
Figure~\ref{fig:steady_states} depicts the steady state solutions for both models for two Deborah numbers and two values of parameter $b$. The nonlinear parameters are set to $\alpha=0.3$ and $\varepsilon=0.3$, respectively.\par

For both models, the velocity and thickness profiles appear to be rather independent of the inlet stress condition. However, the transversal stress component changes significantly close to the die with varying initial conditions. With increasing $x$, the solutions approach each other and seem to be independent of the boundary condition. This was already reported before for the case of fiber spinning of an XPP fluid~\cite{vanderwalt12} and can be understood as a fading memory of the deformation pre-history in the die, which leads to the particular inlet stress condition. This transition occurs within a characteristic time reasonably linked to the relaxation time of the material, which in turn corresponds to a spatial position, as the time within which a fluid element has already experienced deformation can be linked to its position using the velocity profile, i.e., $dt=dx/v$. For $x\gg D\!e$, i.e., exceeding the characteristic time, the pre-history of deformation does not influence the transversal stress any more and is exclusively determined by the deformation outside the die. Note that this effect is not visible in the normal stress difference $\nu_s$ and the axial stress component $t_s$, which are almost independent of the initial stress condition.

\subsection{Refined stress boundary condition}
\label{sec:eff_visc}
\begin{figure}
\centering
\includegraphics[width=1.\textwidth]{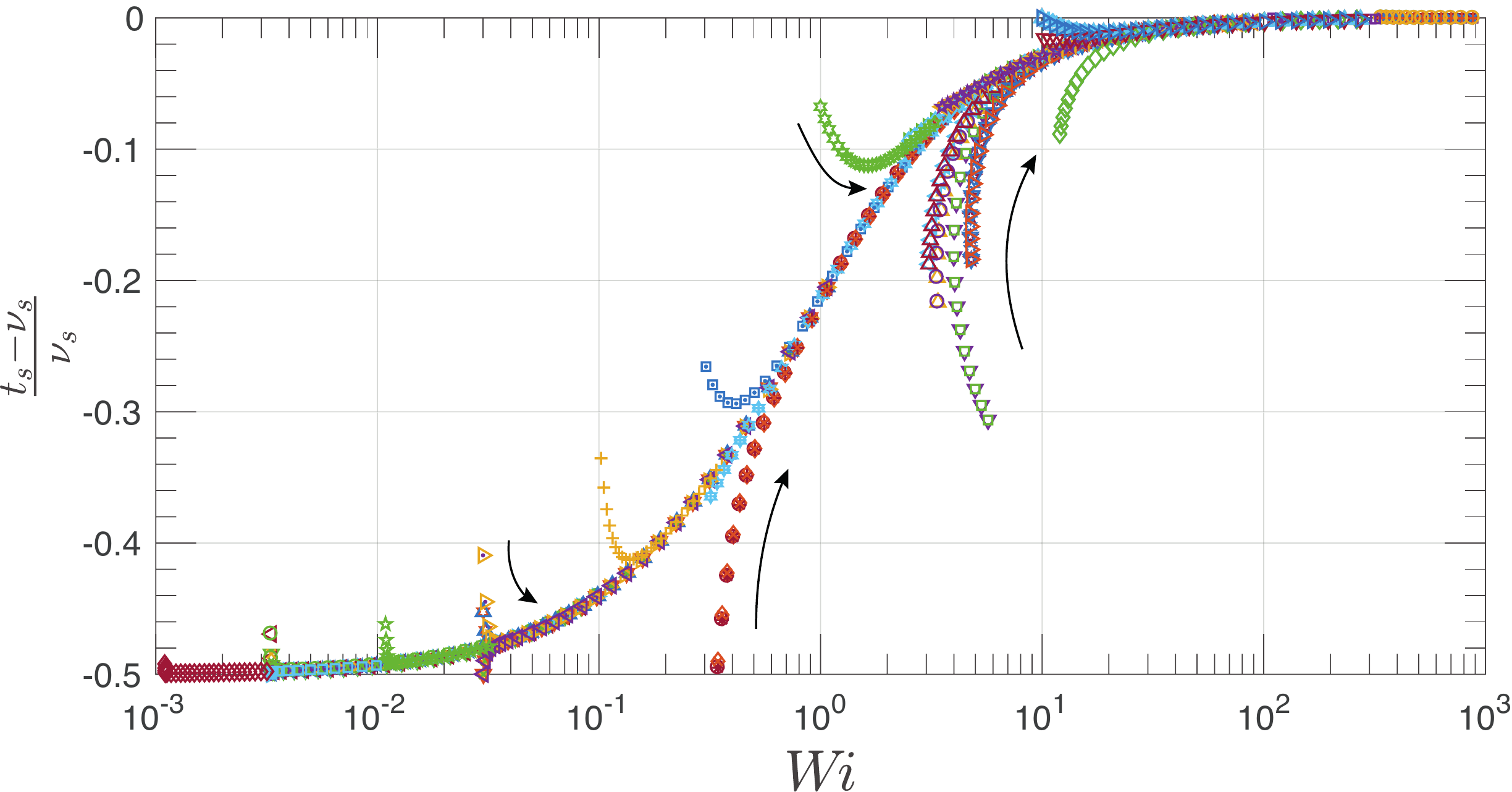}
\caption{Ratio of transversal stress to normal stress difference as a function of the Weissenberg number for the Giesekus model with $\alpha=0.3$, using the empirical boundary condition~(\ref{eq:str_bc_naive}) with various values of $b$. Every data set corresponds to a particular draw ratio and a point-wise evaluation along the $x$-axis. The arrows indicate the increasing $x$-position, i.e., increasing time.}
\label{fig:Giesekus_emp_BC_str}
\end{figure}

Following the idea that there exists a characteristic length of transition after which the deformation is independent of the initial stress condition, we plot the transversal stress profile as a function of the Weissenberg number $W\!i = D\!e\,\partial_x u$, by evaluating both quantities point-wise along $x$. The result is shown by Fig.~\ref{fig:Giesekus_emp_BC_str} for the Giesekus model with $\alpha=0.3$. Every single curve corresponds to a different draw ratio, with various values of $b$ in the inlet stress condition~(\ref{eq:str_bc_naive}) and arrows indicating the direction of increasing $x$-position. It can be seen that with increasing $x$, i.e., with the fluid propagating towards the outlet, all curves collapse to a common master curve, which appears to be independent of the draw ratio. Similar behavior is also observed for the PTT model, including various values of the nonlinear parameter.\par

\begin{figure}
\centering
\includegraphics[width=0.85\textwidth]{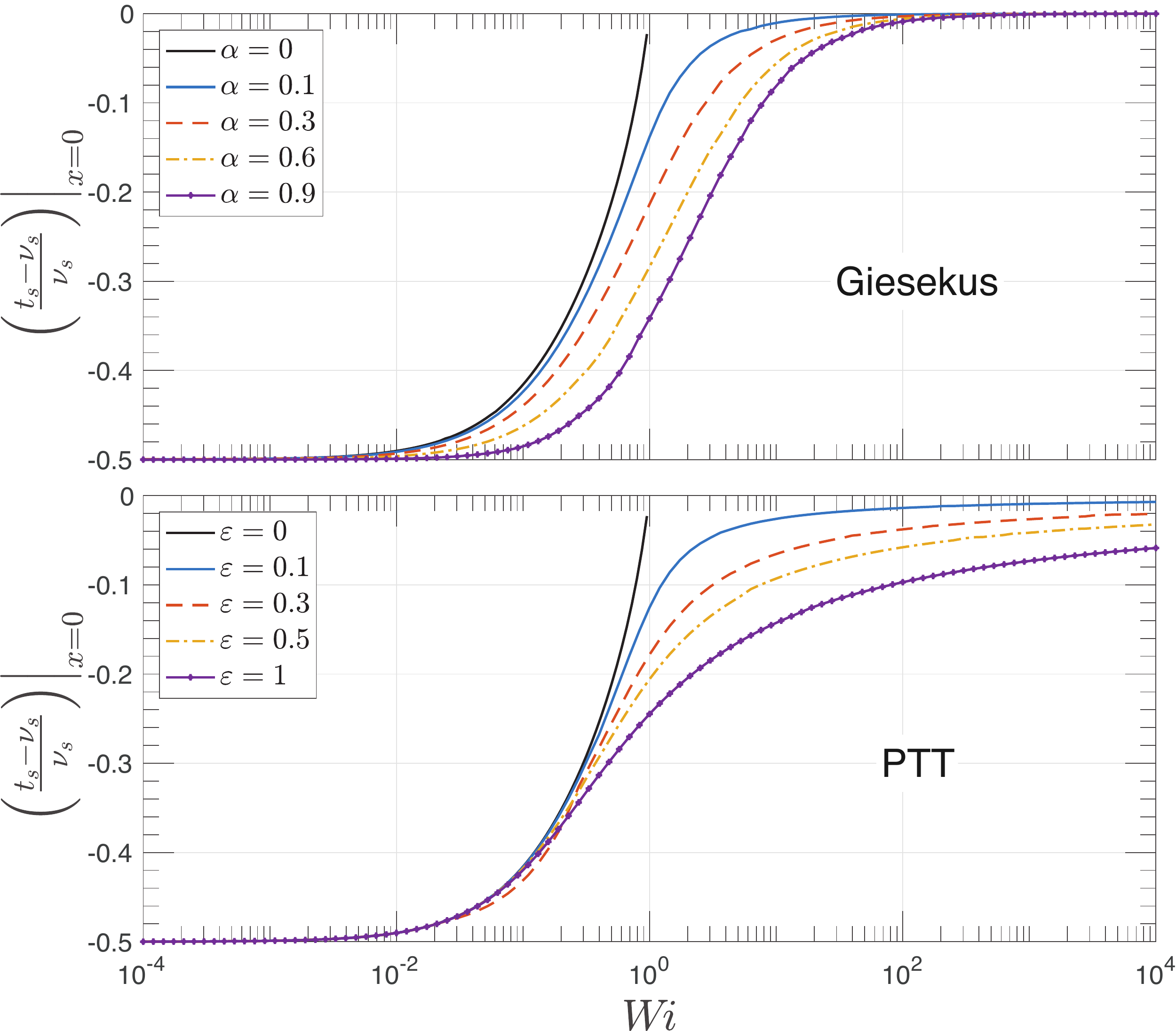}
\caption{Master curves of the transversal stress as a function of the Weissenberg number for the Giesekus and PTT model for several nonlinear parameter values. The curves are obtained by numerical interpolation of the data obtained using initial stress condition~(\ref{eq:str_bc_naive}), after it is truncated by the transition regime close to the inlet. The limit of the UCM model ($\alpha,\varepsilon\rightarrow0$) is not accessible for large Weissenberg numbers, as no steady state solutions exist.}
\label{fig:int_str_all}
\end{figure}

The master curve for a particular model and nonlinear parameter value is obtained by truncating the transition regime, followed by numerical interpolation. Figure~\ref{fig:int_str_all} depicts several of these master curves for the Giesekus and PTT models. Using this master curve, we can now define an initial stress condition based on the physical assumption that the pre-deformation occurring in the die is of the same type as between inlet and outlet, i.e., a planar elongation with varying strain rate. While this assumption may be questionable regarding the different flow properties inside and outside the die, it is still a reasonable one if we want to decouple the effect of die flow as much as possible from the actual casting deformation. In particular, it is identical to the free boundary condition presented by \citet{papanastasiou_96}. Another possibility would be to follow \citet{barborik17}, who used the fully developed flow profile in a rectangular channel to obtain the initial stress condition. This approach, however, increases the computational complexity significantly and introduces additional parameters for the channel. For the rest of this paper, boundary condition~(\ref{eq:str_bc_naive}) is replaced by setting the transversal stress to the value obtained from the interpolated master curve. In general, it should be noted that a reasonable derivation of inlet boundary conditions from the physical setup is a highly non-trivial and yet unsolved problem. Moreover, the stability behavior can depend significantly on the choice of boundary conditions, as for instance demonstrated by \citet{renardy02} in the limit of infinite Deborah number of a UCM model.
\par

\section{Linear stability analysis}
\label{sec:stability}
\subsection{Perturbation equations}
In order to determine the critical parameters beyond which draw resonance occurs, the following ansatz for the variables is posed,
\begin{align}
\label{eq:linstab_ansatz}
\bs{v}(x,t) &= \bs{v_s}(x)+\delta\sum_{n=0}^\infty\left(\bs V_{\!n}(x)\,e^{\omega_n t} + \bs V_{\!n}^*(x)\,e^{\omega_n^* t}\right),
\end{align}
which splits up the state vector $\bs{v} = (u,h,\nu,t)$ into the steady state and a time dependent perturbation part, which is decomposed into perturbation modes with complex perturbation $\bs{V_n} = (u_sU_n,h_sH_n,\nu_sN_n,t_sT_n)$, and corresponding complex eigenvalue $\omega_n=\omega_{n,R} + i\,\omega_{n,I}$. The real part $\omega_{n,R}$ denotes the growth rate and the imaginary part $\omega_{n,I}$ the frequency. The asterisk denotes the complex conjugate. The growth rate determines the stability\footnote{As pointed out by one of the referees, the assumption that the linear stability can be determined from the eigenvalues alone seems to be proven rigorously only in the case of Newtonian fluids~\citep{hagen01} and is therefore assumed \textit{a-priori} in this work.} of the mode, as $\omega_{n,R}>0$ leads to an exponential amplification of an initial perturbation, i.e., unstable behavior, and $\omega_{n,R}<0$ leads to an exponential damping, i.e., stable behavior. We focus on infinitesimal small perturbations only by setting $\delta \ll 1$. Ansatz~(\ref{eq:linstab_ansatz}) is plugged in Eqs~(\ref{eq:1D_cont}), (\ref{eq:1D_mom}) and (\ref{eq:1D_const}), keeping only terms linear in $\delta$. This linearization enables us to decouple the perturbation modes and each mode is determined by the same system of differential equations,
\begin{subequations}
\label{eq:pert_eqs}
\begin{align}
\begin{split}
U' &= \frac{u_s'}{u_s}\left[N-U+\frac{D\!e}{Z_{\nu,s}}\left(\vphantom{\frac{2}{\tilde\mu_s}}\left(\omega H+\left(\omega+3\,u_s'-\nu_s\left(\alpha + \varepsilon \,Z_{\nu,s}\right)\right) N\right)\right.\right.\\
&\,\,\,\, \left.\left. -\,2\,t_s\left(\frac{2}{\tilde\mu_s}-\left(\alpha+\varepsilon\,Z_{\nu,s}\right)\right)T\right)\right],
\end{split}\\
H' &= -\left(\frac{\omega}{u_s}H+U'\right),\\
N' &= -H',\\
\begin{split}
T' &= \frac{1}{D\!e\,u_s\,t_s}\left[t_s\,Z_{xx,s}\,U+2\,u_s\left(1+D\!e\,t_s\right)U'+\varepsilon\,D\!e\,\nu_s\,t_s\,Z_{xx,s}\,N\right.\\
&\,\,\,\, \left. -\left(2\,u_s'+D\!e\,t_s\left(\omega+t_s\left(\alpha+2\,\varepsilon\,Z_{xx,s}\right)\right)\right)T\right],
\end{split}
\end{align}
\end{subequations}
which constitutes a homogeneous eigenvalue problem typical for a linear stability analysis. Note that we omit the index `$n$' from now on for the sake of convenience. The critical draw ratio $D\!r_c$ and the frequency at criticality $\omega_{I,c}$ are determined by the neutral stability condition $\omega_R=0$, where we focus on the most unstable mode, i.e., largest $\omega_R$.\\
The boundary conditions for system (\ref{eq:pert_eqs}) are dictated by conditions~(\ref{eq:bc}) and the prescription of the transversal stress at the inlet. As all these equations are time independent, the corresponding perturbations must vanish at these positions. Additionally, as system~(\ref{eq:pert_eqs}) is homogeneous, we need to specify the amplitude of perturbations. This choice is arbitrary and does not influence the stability predictions. Following previous work \citep{bechert16,bechert17}, we fix the perturbation of the force at the outlet, so that the boundary conditions read
\begin{subequations}
\begin{align}
U(0) = H(0) = t_s(0)T(0)-\nu_s(0)N(0) &= 0,\\
U(1) &= 0,\\
H(1) + N(1) &= 0.1.
\end{align}
\end{subequations}

\subsection{Solution methods}
The perturbation equations are solved by applying two distinct methods. The first consists, analogue to the solution of the steady state equations, in a numerical continuation with \textsc{auto-07p}\citep{auto07p}, but this time with three continuation parameters as the real and imaginary parts of the eigenvalue increase the degrees of freedom by two. The procedure is to fix $\alpha$ and $\varepsilon$ to the desired values and to start with the solution in the Newtonian limit $D\!e\rightarrow0$, for which an analytical expression exists~\citep{renardy06}. The critical conditions are then found by varying $D\!r$, $\omega_R$, and $\omega_I$ simultaneously until $\omega_R=0$. Afterwards, $\omega_R=0$ is fixed and the Deborah number is used as primary continuation parameter, together with $D\!r_c$ and $\omega_{I,c}$ as secondary parameters, following the neutral stability curve in the parameter space.\par

Despite its simple implementation and high computational efficiency, this method has one major drawback. It is tacitly assumed that the most unstable mode of the Newtonian limit remains the most unstable for all configurations, i.e., the frequency at criticality changes continuously under variation of the control parameters. As shown in the following, however, this is not always the case. We therefore determine additionally the eigenvalue spectrum of Eqs.~(\ref{eq:pert_eqs}) to determine possible changes in the most unstable mode. For this purpose, the system is discretized with a Chebyshev collocation method utilizing the \textsc{chebfun} \citep{chebfun} framework for \textsc{matlab} and the eigenvalues and eigenfunctions are computed with the intrinsic solver. In contrast to the method of numerical continuation, the critical conditions have to be found by manually stepping through the parameter space to find the neutral configuration of zero growth rate. Here we use a minimum incremental step size of $0.1$ for the critical draw ratio, leading to an accuracy of $\pm0.05$. As this procedure increases the computational effort and decreases the accuracy, this method is mostly used to verify, and if necessary correct, the results obtained by numerical continuation.\par

\subsection{Giesekus model and UCM limit}
\begin{figure}[t]
\centering
\includegraphics[width=\textwidth]{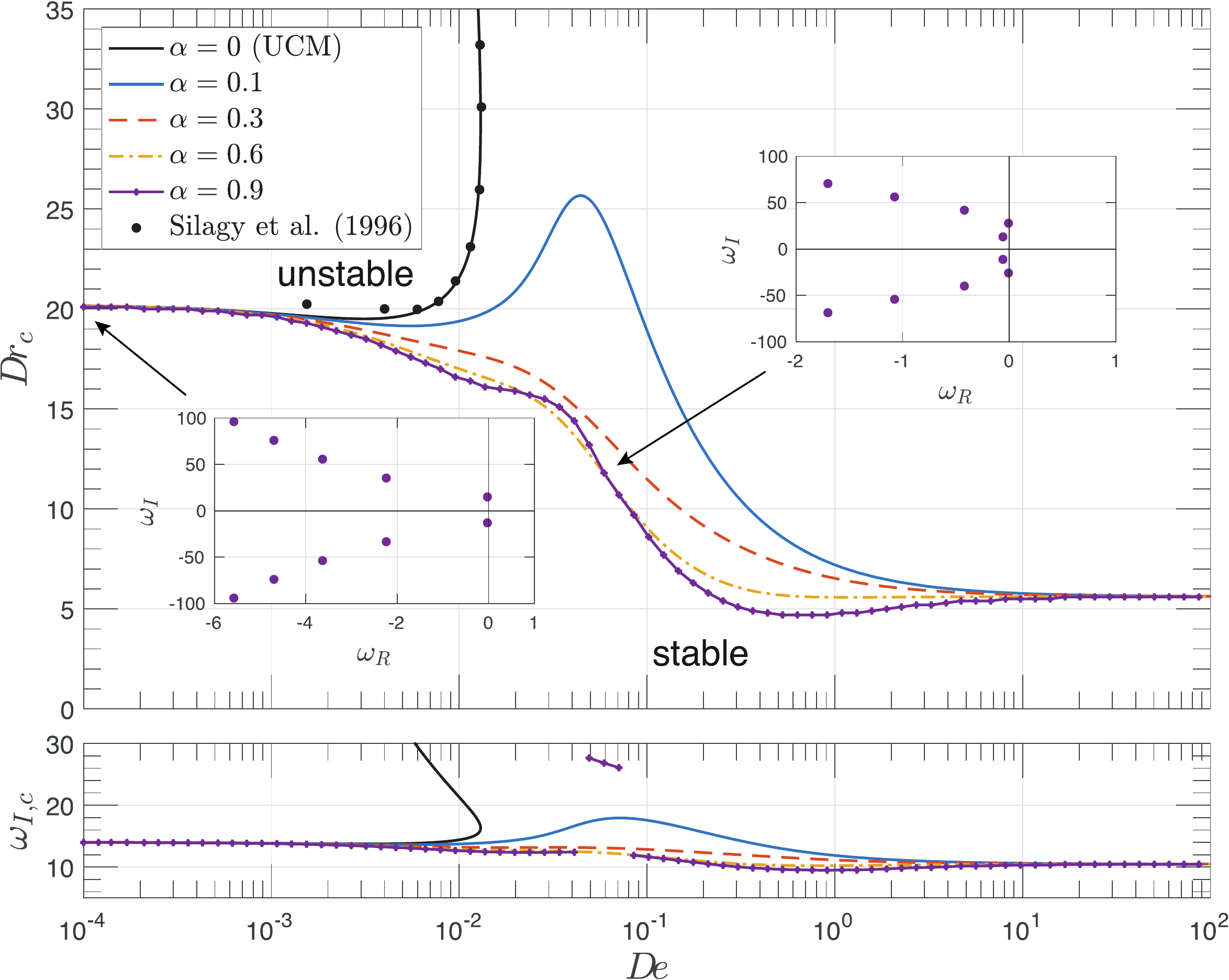}
\caption{Critical draw ratio (top) and frequency at criticality (bottom) as a function of the Deborah number for the Giesekus model for $\alpha=0,\,0.1,\,0.3,\,0.6,\,0.9$, together with comparison to the data of \citet{silagy96} in the UCM limit. The insets show the eigenvalues of the most unstable modes for $D\!e=10^{-4}$ (left) and $D\!e=0.06$ (right) for $\alpha=0.9$.} 
\label{fig:Giesekus_stab}
\end{figure}

As visualized by Fig.\ref{fig:Giesekus_stab}, the Giesekus model predicts a mostly destabilizing effect of increasing elasticity, i.e., increasing Deborah number. For small values of $\alpha\lesssim0.1$, a distinct stability maximum can be observed, while the critical draw ratio decreases monotonically with increasing $D\!e$ for larger values of $\alpha$. The only exception is $\alpha=0.9$, where a small stability minimum around $D\!e\approx 0.5$ is found. In the UCM limit $\alpha\rightarrow0$, which is in good agreement with the results presented earlier by \citet{silagy96}, a largely stabilizing influence of elasticity is present. The curve exhibits a fold point, leading to unconditional stability for $D\!e$ values above this point, and to a second critical draw ratio for $D\!e$ values below this point, above which the system becomes stable again. This second critical draw ratio, however, has never been observed experimentally \citep{hatzikiriakos}, and it is shown below that the drastic increase of $D\!r_c$ can be directly correlated to the unphysical divergence of the elongational viscosity of the UCM model.\par
In particular, the spectral analysis for $\alpha=0.9$ reveals a domain of $D\!e$ within which the second instability mode with higher frequency becomes unstable first. The two insets in Fig.~\ref{fig:Giesekus_stab} showing the first five complex pairs of eigenvalues at different configurations visualize this effect of leading mode switching, which implies a non-continuous jump in the frequency at criticality $\omega_{I,c}$, depicted as well in Fig.~\ref{fig:Giesekus_stab}. Apart from that, the frequency exhibits the same qualitative trend as the critical draw ratio for all values of $\alpha$. In the Newtonian limit of small Deborah numbers, all curves converge to $D\!r_c=20.218$ and $\omega_{I,c}=14.11$, as expected. In the elastic limit of large Deborah numbers, the curves approach again a common value of $D\!r_c=5.6$ and $\omega_{I,c}=10.5$.\par

As can be seen by the right inset in Fig.~\ref{fig:Giesekus_stab}, the real parts of the first two eigenvalues are very close and as a result, the error in $D\!r_c$ introduced by following exclusively the first mode, as done in numerical continuation, is negligible. The jump in frequencies, however, is significant and can be an indicator of a change in the mechanism underlying the instability. As typical values for the Giesekus model for $\alpha$ are rather below $0.5$, we will not further investigate this issue in this work, merely noting that viscoelasticity can lead to switching of the most unstable mode of draw resonance. In general, increasing the nonlinear parameter $\alpha$ appears to have a destabilizing effect, with a small exception for $0.01<D\!e<0.1$, where the neutral stability curves intersect for large values of~$\alpha$.\par

\subsection{PTT model}
\begin{figure}[t!]
\centering
\includegraphics[width=\textwidth]{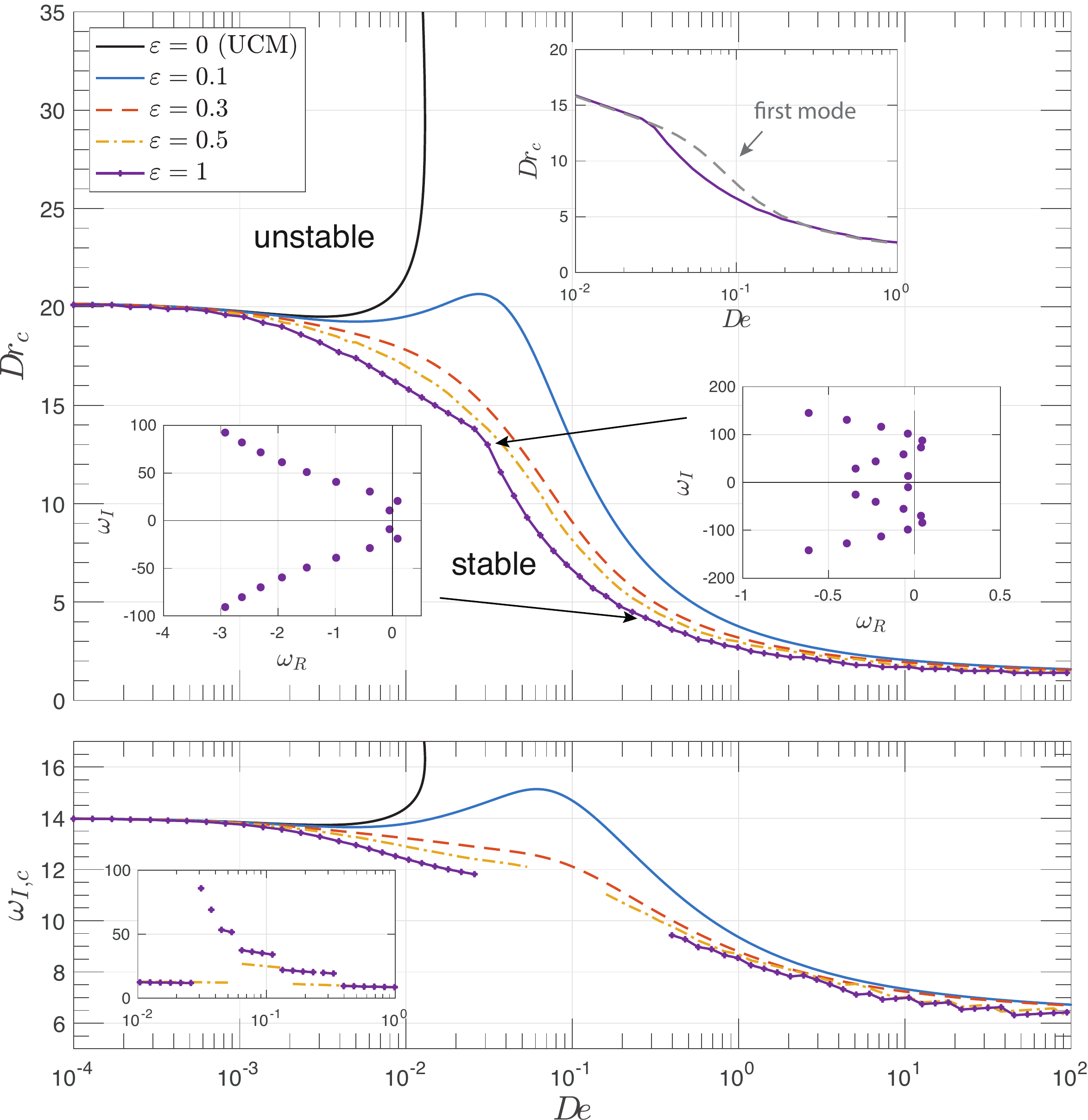}
\caption{Critical draw ratio (top) and frequency at criticality (bottom) as a function of the Deborah number for the PTT model for $\varepsilon=0,\,0.1,\,0.3,\,0.5,\,1$. Two insets show the eigenvalues of the most unstable modes for $D\!e=0.03$ (right bottom) and $D\!e=0.28$ (left) for $\varepsilon=1$. A third inset (right top) compares the critical draw ratio to the prediction following only the first instability mode.}
\label{fig:PTT_stab}
\end{figure}

The stability predictions of the PTT model, as visualized by Fig.~\ref{fig:PTT_stab}, are in qualitative agreement with those of the Giesekus model. Increasing the Deborah number has a primarily destabilizing effect. For small values of the nonlinear parameter $\varepsilon\lesssim0.1$, a stability maximum occurs between $D\!e=0.01$ and $0.1$. In the elastic limit of high Deborah numbers, all neutral stability curves converge to common values $D\!r_c = 1.6$ and $\omega_{I,c} = 6.7$, i.e., lower critical draw ratio and frequency as compared to the Giesekus model predictions. Increasing $\varepsilon$ always leads to destabilization for all Deborah numbers.\par

Similar to the Giesekus model, but more pronounced is the effect of leading mode switching for $\varepsilon\gtrsim0.5$. For $\varepsilon=1$, we find that one of the first six instability modes can become the most unstable one (see bottom insets in Fig.~\ref{fig:PTT_stab} (top)), leading to a change in frequency up to a factor of $7$, which is also shown by the inset of Fig.~\ref{fig:PTT_stab} (bottom). In contrast to the Giesekus model, this mode switching also leads to a significant change in the critical draw ratio, as depicted by the top inset in Fig.~\ref{fig:PTT_stab} (top). \citet{christodoulou00} also observed switching of the most unstable instability mode in their analysis of a multi-relaxation-mode PTT model.\par

\section{Generalized Newtonian fluid models}
\label{sec:GNF}
\subsection{Effective elongational viscosity}
Generalized Newtonain fluid (GNF) models can offer an efficient way to cover the main aspects of viscoelastic fluids, if (non-Newtonian) viscous effects are dominating and purely elastic effects are negligible. In this case, the viscosity is assumed to be a function of the flow properties, mostly the deformation rate. Instead of the shear viscosity $\eta$, we use here the steady effective elongational viscosity $\tilde\mu_s$ as defined by Eq.~(\ref{eq:eff_visc}), as film casting is dominated by elongational deformation.\par

We evaluate $\tilde\mu_s$ for the Giesekus and PTT models for various values of $D\!r$, $D\!e$, and the nonlinear parameters. Analogue to the transversal stress component shown by Fig.~\ref{fig:int_str_all}, the effective elongational viscosity appears to be independent of the draw ratio and different configurations yield parts of a common master curve, which can be interpolated numerically. Using the empirical boundary condition~(\ref{eq:str_bc_naive}), the boundary layer effects close to the die, similar to the ones found for the transversal stress (Fig.~\ref{fig:Giesekus_emp_BC_str}), can be observed. These effects disappear if the interpolated master curve is used as initial stress boundary condition instead.\par

\begin{figure}[t]
\centering
\includegraphics[width=\textwidth]{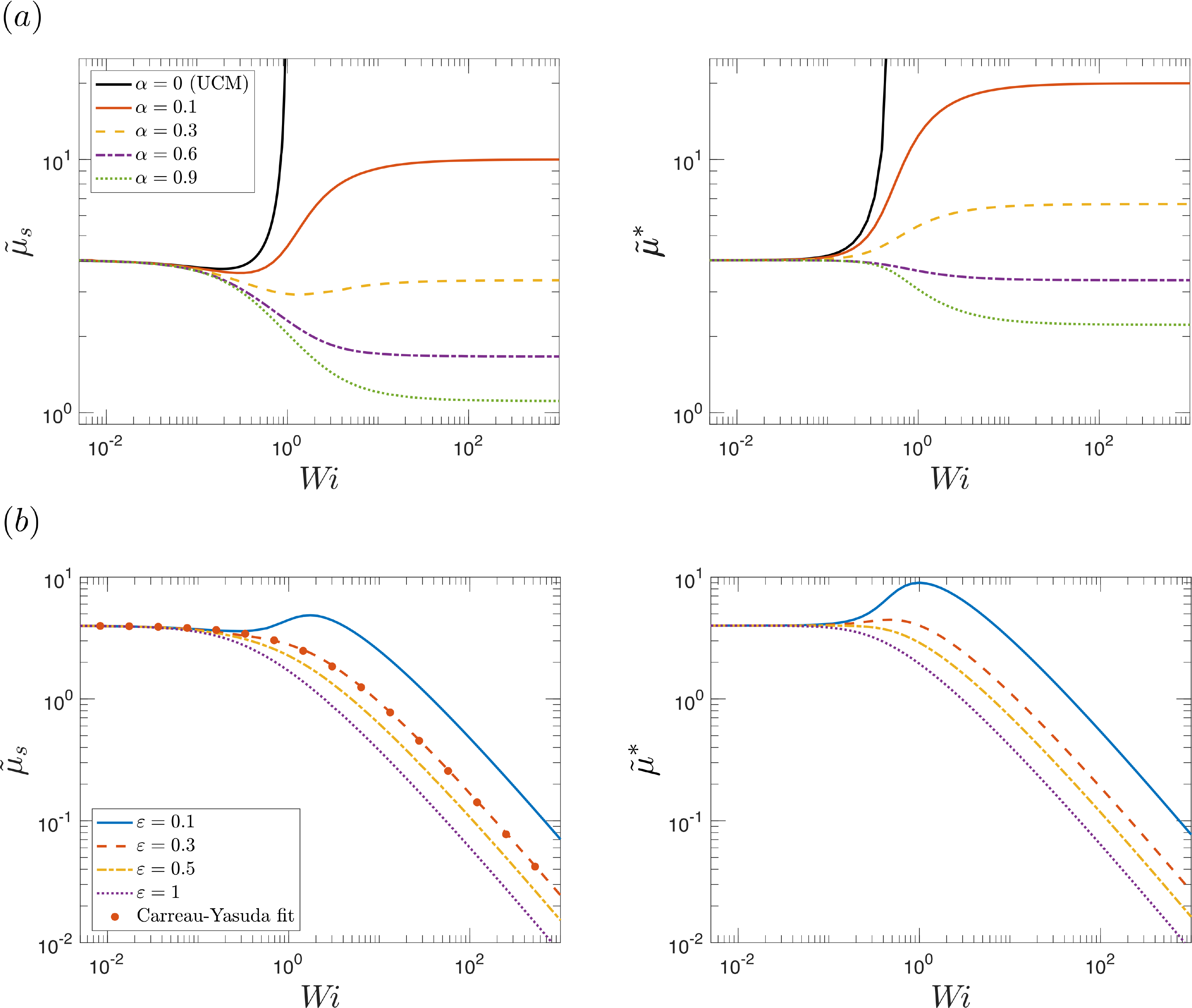}
\caption{Effective viscosities at steady state $\tilde\mu_s$ for the (a) Giesekus and (b) PTT models. In both cases, the quantity appears to depend solely on the Weissenberg number~$W\!i$ and the shapes of the curves resemble the planar elongational viscosities at constant strain rate, $\tilde \mu^*$, which are depicted as well for comparison. For the PTT model~($\varepsilon=0.3$), an empirical fit utilizing Eq.~(\ref{eq:CY}) is indicated.}
\label{fig:eff_visc}
\end{figure}

Figure~\ref{fig:eff_visc} depicts the resulting effective elongational viscosities for both models as a function of $W\!i$, evaluated using $\partial_xu_s$, and compares them to the steady planar elongational viscosity at constant strain rate, $\tilde\mu^*$. Note that it may help here to keep in mind that $W\!i$ can be regarded as a dimensionless strain rate. For all three models, the shape of $\tilde\mu_s$ is very similar to $\tilde\mu^*$. There is, however, a quantitative difference, which is crucial for the draw resonance behavior as shown below. In general it can be stated that the strain hardening effect is diminished as compared to the constant strain rate deformation. For the Giesekus model, all plateau values for large Weissenberg numbers are lowered and a significant strain hardening is present only for small values of $\alpha$. Similarly, the strain hardening maximum predicted by the PTT model becomes less pronounced and vanishes for instance already for $\varepsilon=0.3$, being still present for $\tilde\mu^*$.\par

Comparing now the effective elongational viscosity curves with the neutral stability curves shown in Figs.~\ref{fig:Giesekus_stab} and~\ref{fig:PTT_stab} reveals strong similarities. It is striking that a stability maximum is present if and only if the corresponding effective elongational viscosity exhibits strain hardening. Note that this is not true for the planar viscosity at constant strain rate, which for instance shows strain hardening as well for $\alpha=0.3$ or, respectively, $\varepsilon=0.3$, where increasing $D\!e$ has a purely destabilizing effect. In the UCM limit also shown in Fig.~\ref{fig:eff_visc}(a), the excessive strain hardening clearly resembles the highly stabilizing influence of increasing elasticity.\par

These observations strongly suggest that the effective elongational viscosity plays a key role in the dynamical behavior and in particular in the stability of the process. For this reason, we will analyze in the following the steady state and instability behavior of a GNF model based on the effective elongational viscosity and compare the results to the full viscoelastic model prediction. For the sake of compactness, we focus primarily on the PTT model and refer to the Giesekus model merely in a qualitative way at the end of the section.

\subsection{Reproducibility using a GNF model}
In absence of strain hardening, it is possible to fit the effective elongational viscosity of the PTT model with the expression of the Carreau-Yasuda (CY) model \cite{macosko}, which is given by
\begin{align}
\label{eq:CY}
\tilde\mu_{\rm cy}(W\!i) = \frac{4}{\left[1+\left(k\,W\!i\right)^m\right]^\frac{1-n}{m}}.
\end{align}
For $\varepsilon=0.3$, the free parameters are determined as $k=0.433$, $n = 0.162$, and $m=0.889$ and the corresponding fit is shown in Fig.~\ref{fig:eff_visc}(b). While the effective viscosity is originally obtained exclusively from the steady state solution, the main assumption of the GNF model now is that this viscosity is valid as well for time dependent dynamics. The GNF model is thus given by Eqs.~(\ref{eq:1D_cont}) and (\ref{eq:1D_mom}), supplemented by
\begin{align}
\nu = \tilde\mu_{\rm cy}(D\!e\,\partial_xu)\,\partial_xu.
\end{align}\par
Due to the particular structure of Eq.~(\ref{eq:CY}), it is impossible to analytically solve the GNF model equation for $\partial_x u$, as it occurs with various exponents in the model equations. For this reason, we use $\partial_x u$ instead of $\nu$ as a variable, which leads to the following steady state equations,
\begin{subequations}
\label{eq:CY_steady}
\begin{align}
u_s'' &= \frac{u_s'^2\left(1+k\,D\!e\,u_s'\right)^m}{u_s\,c},\\
h_s' &= -\frac{h_s}{u_s}u_s',
\end{align}
\end{subequations}
with the auxiliary variable
\begin{align}
c = \left(1+n\left(k\,D\!e\,u_s'\right)\right)^m.
\end{align}
The perturbation equations read
\begin{subequations}
\label{eq:CY_pert}
\begin{align}
\begin{split}
U'' &= \frac{1}{u_s^2 c^2}\left[m(1-n)\left(k\,D\!e\,u_s'\right)^m u_s'^2 U + \omega\,c\left(1+\left(k\,D\!e\,u_s'\right)^m\right)u_s'H\right.\\
&\ \ \ \left. +\ (1-n)\left(k\,D\!e\,u_s'\right)^m\left(m+2\,c\right)u_su_s'U'\right],
\end{split}\\
H' &= -\left(\omega\,\frac{H}{u_s}+U\right).
\end{align}
\end{subequations}
\begin{figure}[t]
\centering
\includegraphics[width=\textwidth]{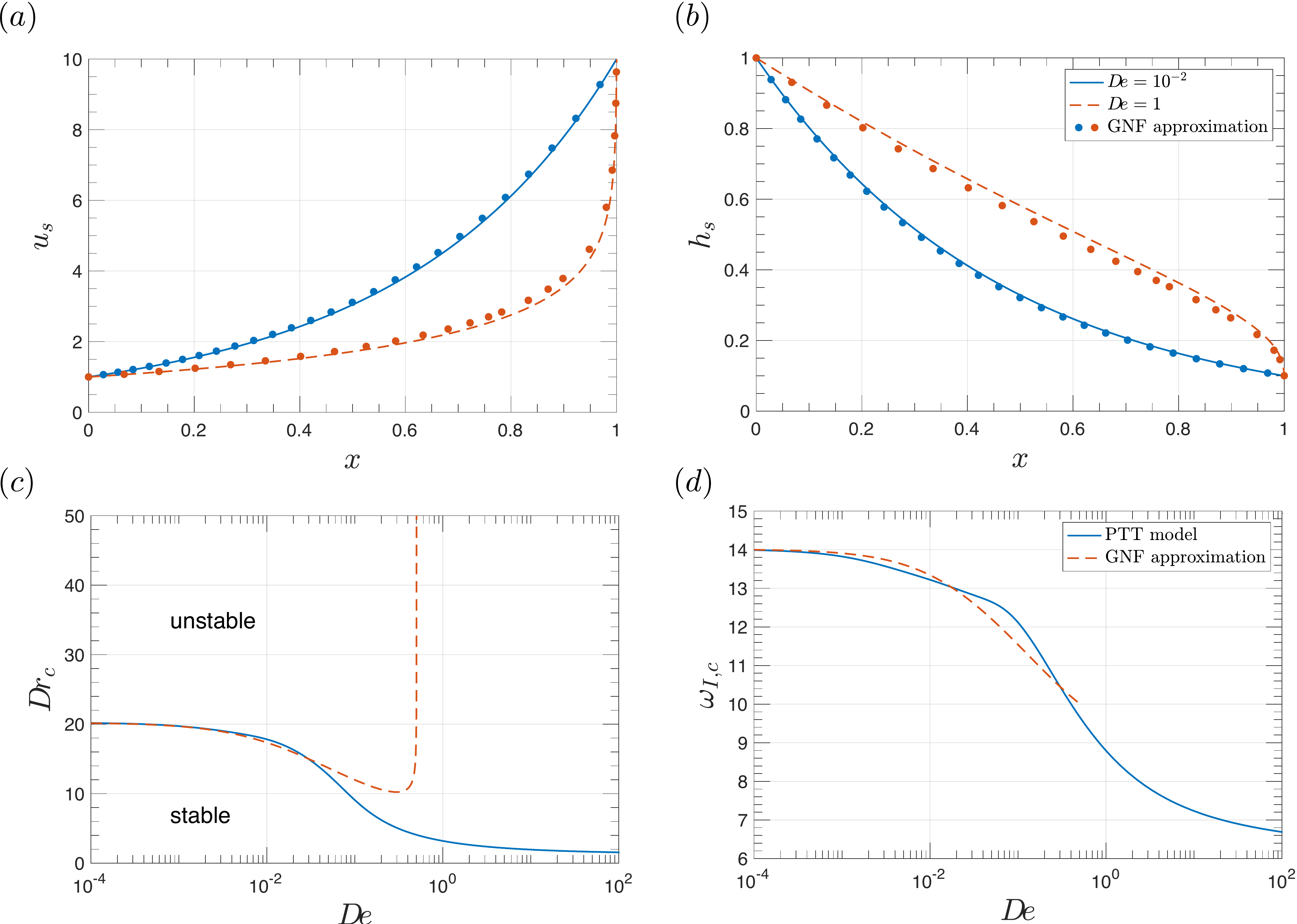}
\caption{Comparison between the PTT model, $\varepsilon=0.3$, and its GNF approximation based on the effective elongational viscosity, which is modeled by the CY Eq.~(\ref{eq:CY}). (a,b) Steady state solutions for axial velocity and film thickness for $D\!e = 10^{-2}$ and $1$ and $D\!r=10$. (c)~Critical draw ratio. (d) Frequency at criticality.}
\label{fig:PTT_GNF}
\end{figure}\par
Figures~\ref{fig:PTT_GNF}(a,b) compare the steady states of axial velocity and film thickness as obtained from this GNF approximation to the results of the PTT model for two values of the Deborah number and $D\!r=10$, showing a high overall agreement of both models. Small deviations are most likely caused by imperfections in the fitting of the CY~model. The neutral stability curves (Fig.~\ref{fig:PTT_GNF}(c)), however, coincide only for $D\!e<0.1$. For larger Deborah numbers, the GNF approximation overpredicts the critical draw ratio and finally diverges at $D\!e\approx0.5$, while the PTT model predicts a monotonically decreasing stability with increasing $D\!e$. Note that the divergent behavior is not caused by a switch of the most unstable mode, as it occurs for the Giesekus and PTT models as discussed above, as higher modes of instability were checked separately. Moreover, the predicted frequency at criticality of the GNF approximation is in good coincidence with the PTT model result until the point of divergence (see Fig.~\ref{fig:PTT_GNF}(d)).\par
This reveals that the use of GNF models for studying the stability is limited at least to rather low Deborah numbers. On the other hand, in this regime the effective elongational viscosity appears to have a dominant influence on the stabiblity behavior. A likely reason for the limitation to low $D\!e$ is the missing explicit time dependence of the stress tensor, which appears in the viscoelastic models applied here in the form of the upper-convected time derivative. As a consequence, there exists an additional, yet unexplored mechanism underlying the stability behavior of systems dominated by elastic effects. The presence of such a mechanism is also visible in the stability results of the Giesekus model~(Fig.~\ref{fig:Giesekus_stab}): While the effective elongational viscosity reaches a plateau value in the elastic limit of large $W\!i$ (Fig.~\ref{fig:eff_visc}(a)), the corresponding critical draw ratio is clearly below the Newtonian limit value of $D\!r_c = 20.218$.\par

\section{Influence of strain hardening and strain thinning}
\label{sec:mech}
Besides this purely elastic effect, the mechanism underlying the (de-)stabilization of non-Newtonian, viscous properties is still unclear. In particular, the question arises why a purely strain thinning material leads to a non-monotonous stability behavior, including a diverging critical draw ratio, as shown in Fig.~\ref{fig:PTT_GNF}(c)). There exists a supposition for a viscous mechanism underlying draw resonance, which is based on the work of \citet{kim96}, and which was refined later~\citep{scheid09,bechert16,bechert17}. The general idea is to follow the propagation of an initial perturbation of the tension $f = h\,\nu$. According to Eq.~(\ref{eq:1D_mom}), this tension is constant along $x$. Evaluation of the continuity equation~(\ref{eq:1D_cont}) at $x=0$, where $h$ is time independent and equal to unity, enables us to write
\begin{align}
\label{eq:mech_1}
\partial_xh|_0 = -\frac{f}{\tilde \mu|_0},
\end{align}
where the GNF model expression $\nu = \tilde\mu\,\partial_x u$ for the normal stress difference was used. This equation reveals that a perturbation of the tension, e.g., at the outlet, instantaneously leads to a perturbation in thickness at the inlet, which then travels downwards, finally causing another perturbation at the outlet and closing the loop of the feedback mechanism. Exploiting the fact that $f$ is constant in space, we can rewrite Eq.~(\ref{eq:mech_1}) to better visualize the relation between perturbations at in- and outlet:
\begin{align}
\label{eq:mech_2}
\partial_xh|_0 = -\frac{\tilde\mu|_1}{\tilde\mu|_0}\,h|_1\,\partial_xu|_1.
\end{align}
\begin{figure}[h]
\centering
\includegraphics[width=\textwidth]{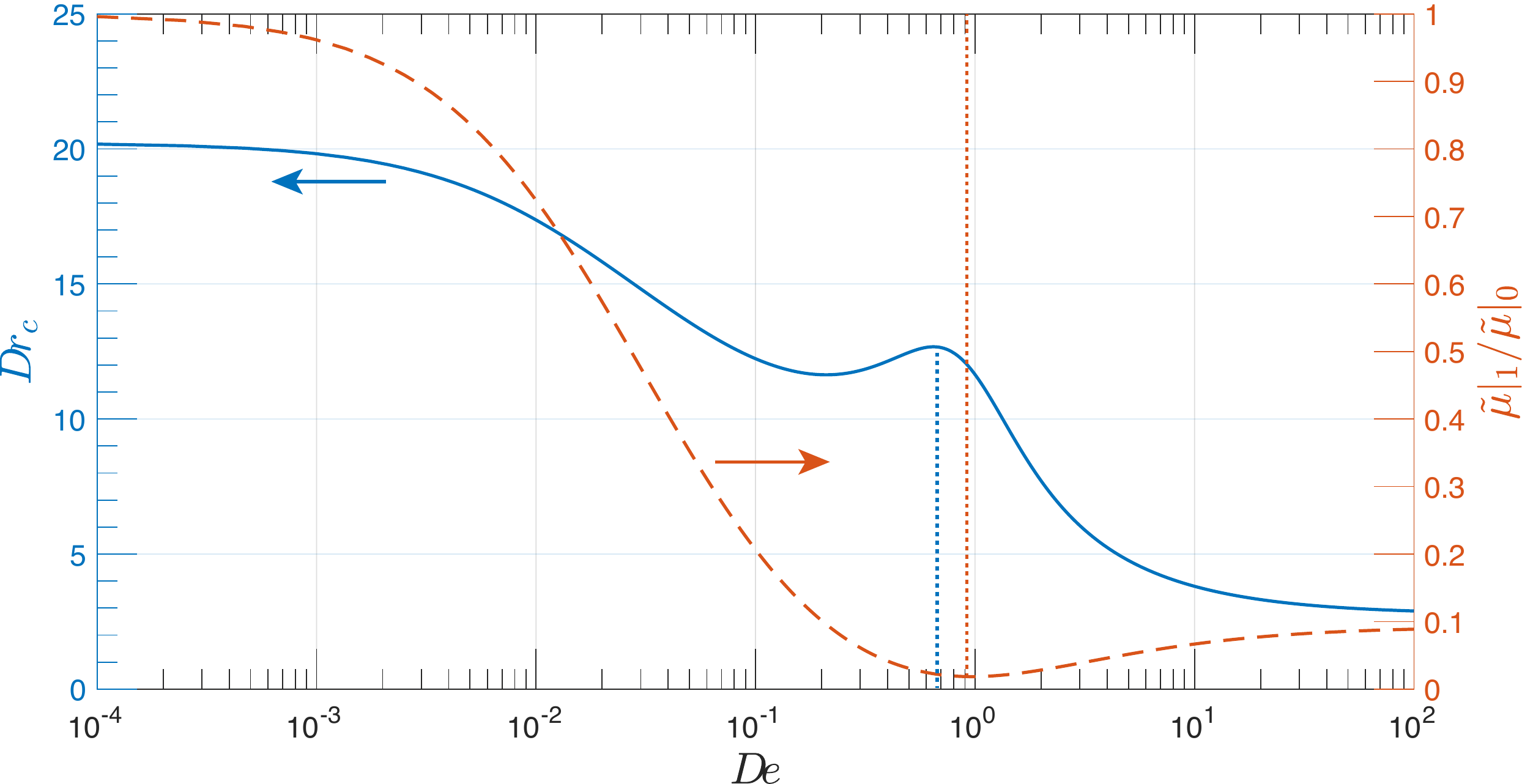}
\caption{Critical draw ratio of the CY~model for $n=0.3$, $m=1$, and $k=1$ (blue, solid), together with the ratio of effective elongational viscosities at outlet to inlet (orange, dashed). The dotted lines indicate the extrema of the two curves.}
\label{fig:CY_stab}
\end{figure}
\par
Further analysis of the CY~model reveals that the neutral stability curve infact does not always diverge, but exhibits for some parameter combinations merely a stability maximum with an overall destabilizing trend for increasing Deborah number. An example is shown in Fig.~\ref{fig:CY_stab}, where $m=k=1$ and $n=0.3$, and where a stability maximum occurs close to $D\!e=1$. Note that $n$ tunes the slope of the power-law regime for $k\,W\!i\gg1$, while $m$ sets the sharpness of the transition from Newtonian to power-law behavior. The shown case here basically differs from the one fitted to the PTT viscosity above in the value of $n$, leading to a less steep power-law slope.\par

Following Eq.~(\ref{eq:mech_2}), we now look at the ratio of effective elongational viscosities at outlet to inlet, as this is the quantity which differs from the Newtonian case. It is plotted in Fig.~\ref{fig:CY_stab} as well and exhibits a local minimum close to the stability maximum. With the viscous stability mechanism described above in mind, Eq.~(\ref{eq:mech_2}) can be interpreted in the following way. While a perturbation at the outlet leads to a perturbation at the inlet, this effect is scaled by the ratio of effective viscosities. This seems to be reasonable as the mechanism is based on the perturbation of tension, which is proportional to the viscosity. Therefore, a perturbation at the outlet has less effect on the perturbation at the inlet, if the viscosity at the inlet is larger as compared to the outlet. If the ratio of viscosities is below a threshold, the instability is completely suppressed and the critical draw ratio diverges, as observed in Fig.~\ref{fig:PTT_GNF}(c). Note that a similar argumentation was already successfully used in the past to explain the change in stability behavior caused by the neck-in in film casting~\citep{bechert16}.\par

As we are employing merely the steady states of the elongational viscosities in this reasoning, we call this mechanism the ``static mechanism'' underlying non-Newtonian draw resonance. However, according to this explanation one would expect a purely stabilizing effect of a strain thinning fluid like the analyzed one, as the strain rate is expected to increase along $x$ during the casting and the thinning effect is more pronounced for larger $D\!e$. Instead, the overall trend appears to be destabilizing with increasing $D\!e$ and it is known that pure power-law models lead to an increase/decrease of $D\!r_c$ for strain hardening/thinning materials \cite{aird83}. For this reason, there has to exist a second mechanism, which we will call the ``dynamic mechanism''.\par
\begin{figure}[t]
\centering
\includegraphics[width=0.6\textwidth]{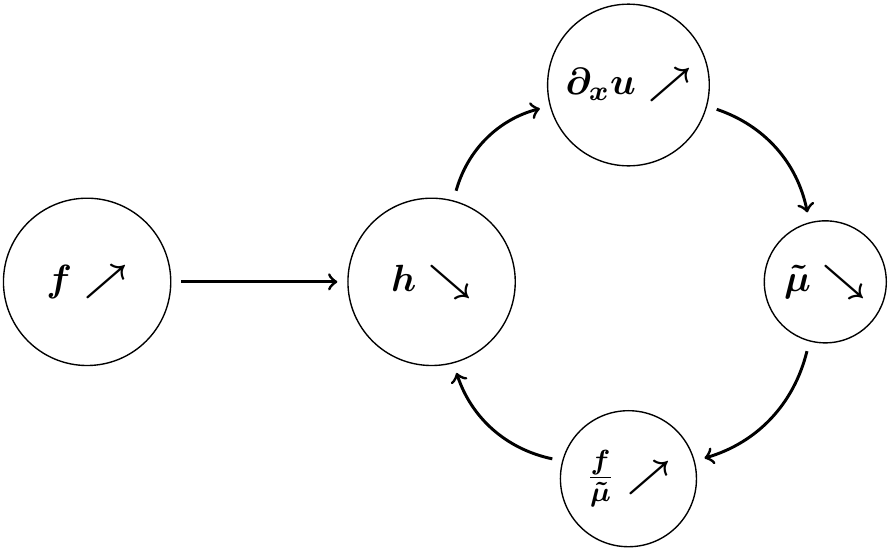}
\caption{Visualization of the dynamic mechanism underlying the destabilizing effect of strain thinning. While $f$ is constant in space, all other quantities correspond to a position close to the inlet.}
\label{fig:mech}
\end{figure}
The dynamic mechanism is visualized in Fig.~\ref{fig:mech} for a strain thinning material and can be understood as follows. According to Eq.~(\ref{eq:mech_1}), a perturbation of the tension, e.g., a temporary increase, leads to a stronger (negative) gradient of the film thickness at the inlet and therefore to a smaller film thickness close to the inlet. Due to continuity, this increases the velocity and thus the strain rate close to the inlet as well. For a strain thinning/hardening material, this has the consequence of a decreasing/increasing effective viscosity, which in turn leads to an increase/decrease of the ratio on the right-hand side of Eq.~(\ref{eq:mech_1}), which finally amplifies/damps the initial effect. Therefore, the dynamic mechanism leads to a destabilization/stabilization of strain thinning/hardening, while the static mechanism has an opposing effect.\par

Whether the static or the dynamic mechanism is more dominant and governs the stability behavior depends on the particular shape of the effective elongational viscosity. For pure power-law behavior, the dynamic mechanism is clearly stronger, but in the case of the CY~model, the situation is more subtle. As long as the ratio of outlet to inlet effective viscosity does not deviate too much from unity, the dynamic mechanism is more pronounced than the static. If, however, the material at the inlet is rather governed by the Newtonian plateau regime, while the material at the outlet enters already the power-law regime, the non-Newtonian effect close to the inlet and thus the dynamic mechanism is weak compared to the static mechanism, and the stability behavior changes qualitatively.\par

It should be noted that it was separately verified that a time-independent GNF model, which depends on the steady state of the strain rate only, yields indeed a purely stabilizing/destabilizing effect for strain thinning/hardening materials, as the dynamic mechanism is absent here. A similar effect was observed by \citet{scheid09}, when they found that cooling can have a destabilizing effect if the Stanton number is very large so that the material temperature is actually completely governed by the ambient temperature and therefore time independent.

\section{Conclusions and Outlook}
We studied the influence of viscoelastic, and in particular non-Newtonian, effects on the draw resonance instability in film casting. For this purpose, a comprehensive framework for a linear stability analysis of the Giesekus and PTT models, with the UCM model as limit case, was derived. The boundary condition for the initial stress was chosen in a way that a deformation history caused by a particular flow inside the die is excluded from the analysis, which is consistent with the free boundary condition method of \citet{papanastasiou_96}. Both the Giesekus and the PTT models show an overall destabilizing trend, except for the UCM limit, where a strong stabilizing effect of elasticity is present. This qualitative difference could be related to the unphysical, diverging elongational viscosity of the UCM model. For this reason, the UCM model is highly not recommended for studies of viscoelastic film casting or related processes like fiber spinning. For large values of the nonlinear parameters of the Giesekus and PTT models, a switch of the most unstable mode, and thus a discrete change of the frequency at criticality can be observed.\par

While it is possible for low values of the Deborah number to reproduce the results of the viscoelastic models using GNF models based on the effective elongational viscosity, which was introduced as a key quantity for the analysis, this approximation fails if elastic effects become too pronounced. Nevertheless, the GNF models, in particular the CY model, enable us to shed light on the two opposing mechanisms underlying the effects of strain hardening and thinning materials on the stability.\par

Apart from new insights in non-Newtonian effects in draw resonance, the present results strongly indicate the existence of an additional, purely elastic stability mechanism. Besides the failure of the GNF approximation for high Deborah numbers, a change in the basic mechanism can also be indicated by the observed switching of the most unstable instability mode. Given that the common approaches to explain draw resonance are based on a viscous material, the question arises whether we can still speak of draw resonance in the high elastic regime, or whether this is another type of instability. This is similar to the transition from draw resonance to the Rayleigh-Plateau instability when surface tension in fiber spinning is increased \citep{bechert17}. In a weakly nonlinear stability analysis, \citet{gupta15} found a transition from a supercritical to a subcritical instability if the Deborah number exceeds a threshold value, which is another hint for this hypothesis. However, it has to be noted that this analysis is at least partially wrong, as pointed out by \citet{gallaire16}. Further work on the nature of the elastic instability is therefore needed, and it is hoped that the present study can serve hereby as a solid basis.

\section*{Acknowledgments}
The author thanks Manuel Alves, Fran\c cois Gallaire, Rob Poole, and Benoit Scheid for fruitful discussions and useful tips.

\bibliographystyle{elsarticle-num-names}

\end{document}